\documentclass[aps,pra,twocolumn,superscriptaddress,floatfix,showpacs]{revtex4}

\usepackage[usenames]{color}

\newcommand{\dens}[1]{{\rm [#1]}}

\newcommand{\tHe}{\mbox{$^3$He}}
\newcommand{\fHe}{\mbox{$^4$He}}

\newcommand{\tkSR}{\,^{3}k_{\rm SR}}
\newcommand{\fkSR}{\,^{4}k_{\rm SR}}
\newcommand{\tgA}{^3\gamma_{\rm A}}
\newcommand{\fgA}{^4\gamma_{\rm A}}

\newcommand{\boldbox}[1]{\mbox{\boldmath $#1$}}

\def\be{\begin{eqnarray}}
\def\ee{\end{eqnarray}}

\usepackage{graphicx}

\begin{document}


\newcommand{\abs}{}

 \title{Method for deducing anisotropic spin-exchange rates}

\author{T. G. Walker}
\affiliation{Department of Physics, University of Wisconsin-Madison,
		Madison, WI 53706 }
\author{I. A. Nelson}
\affiliation{Columbia--St. Mary's Hospital, Milwaukee, WI 53211}
\author{S. Kadlecek}
\affiliation{Department of Radiology, University of Pennsylvania, Philadelphia, PA 19104}
\date{\today}

\begin{abstract}{Using measured spin-transfer rates from alkali atoms to \tHe, combined with spin-relaxation rates of the alkali
atoms due to
\tHe\ and \fHe, it should be possible to differentiate between isotropic and anisotropic spin-exchange.  This would give a fundamental limit on the $^3$He polarization attainable in spin-exchange optical pumping.  For K-He, we find the limit to be 0.90$\pm$0.11. }\end{abstract}


\maketitle

\section{Introduction and Motivation}

The spin-dependent interactions governing spin-exchange collisions between alkali-metal atoms and noble-gas atoms are \cite{WalkerRMP}
\be
V=\alpha(R) {\bf S}\cdot{\bf K}+\beta(R)(3{\bf S}\cdot{\bf \hat R \hat R\cdot K-S\cdot K})
\ee
where $\alpha$ is the strength of the Fermi-contact or isotropic hyperfine interaction between the alkali-metal electron spin $\bf S$ and the noble gas nuclear spin $\bf K$, and $\beta$ is the strength of the anisotropic hyperfine interaction.  Both $\alpha$ and $\beta$ depend on the interatomic separation $\bf R$.

Anisotropic spin-exchange was recently considered by Walter {\it et al.}\cite{Walter98} and, on the basis of theoretical arguments that have generally been successful in explaining the size of various alkali--noble-gas spin-interactions, was found to be a small effect. 
If present, anisotropic spin-exchange would modify the dynamics of polarization transfer from an alkali vapor of number density \dens{A} to the helium nucleus  to
\be
{dP_{\rm He}\over dt}&=&k_\alpha\dens{A}\left(P_{\rm A}-P_{\rm He}\right)+\nonumber\\
&&k_\beta\dens{A}\left(-{P_{\rm A}\over 2}-P_{\rm He}\right)-\Gamma_w  P_{\rm He} \label{rateeqn}
\ee
where $k_\alpha$ and $k_\beta$ are the rate coefficients arising from the two interactions and $\Gamma_w$ represents depolarization at the wall of the gas enclosure. ÊNote that anisotropic spin-exchange tends to polarize the He nuclei in the direction opposite that of the alkali polarization.  In the presence of completely polarized alkali vapor and non-relaxing walls, nearly achievable in practice, the anisotropic interaction would limit the maximum attainable polarization to
\be
P_{\rm max}={k_\alpha-k_\beta/2\over k_\alpha+k_\beta}\label{plimit}
\ee
Walter {\it et al.} \cite{Walter98}
predicted $P_{\rm max}=0.96$ for Rb-\tHe\ and $0.95$ for K-\tHe.  Extensive experiments at Wisconsin and NIST \cite{Chann02c,Chann03,Babcock06}  have shown that some unknown spin-relaxation mechanism limits the \tHe\ polarization, even under supposedly ideal
conditions, to less than 80\% for both Rb and K-Rb mixtures.  Could one source of this relaxation be anisotropic spin-exchange?  Here
we present a method for experimentally answering  this question, by
 deducing $k_\beta$ from spin-exchange and alkali-metal spin-relaxation measurements.

\section{Limits from Wall Relaxation Studies}

The approach of $P_{\rm He}$ to saturation in the presence of a polarized alkali vapor can be experimentally characterized by its saturation level $P_{{\rm He}}^\infty$ and rate of approach to saturation $\Gamma$. ÊFrom Eq.~\ref{rateeqn},
\be
	P_{\rm He}^\infty &=& P_{\rm A}{(k_\alpha-k_\beta/2)[A]\over \Gamma_w+(k_\alpha+k_\beta)[A]}\\
	\Gamma &=& \Gamma_w+(k_\alpha+k_\beta)[A] 
\ee
For any given measurement $P_{\rm He}(t)$ performed at constant $P_{\rm A}\dens{A}$, $\Gamma_w$ can be eliminated, leaving
\be 
	k_\alpha-k_\beta/2={P_{\rm He}^\infty\Gamma\over \dens{A}P_{\rm A}} 
\ee
The quantity $k_{\rm SE}=k_\alpha-k_\beta/2$ is what is observed in recent spin-exchange measurements\cite{Baranga98c,Chann02c,Babcock06}.
It might appear that measurements of $P_{\rm He}^\infty$ or $\Gamma$ as a function of \dens{A} would allow determination of $k_\alpha + k_\beta$.  But it is now well-established \cite{Babcock06,Chann02c} that $\Gamma_w$ depends strongly on \dens{A}, making this approach not feasible.  

The latest wall  studies \cite{Babcock06}, surveying many cells having a range of surface to volume ratios $S/V$, found that  the observed polarization is well-described by
\be
{P_{\rm He}^\infty\over P_{\rm A}}={1\over 1+X}
\ee
where $X$ is  of the form 
\begin{equation}
X=X_0+X_1{S\over V}
\end{equation}
If we assume that $X_0$ represents the fundamental (wall-independent) effects of the anisotropic hyperfine interaction, comparison to Eq.~\ref{plimit} yields
\be	
X_0 = {3k_\beta\over 2k_\alpha-k_\beta }
\ee
The factor $X_0$, which would represent a limit on $P_{\rm He}$ from collisions in the gas,  could be as small as 0 and as large as $0.15$ \cite{Babcock06}.  ÊVariability in measured $X_1$ limits the certainty of the results, presumably due to its very sensitive dependence on the exact physical and chemical nature of the wall.

\section{Method}
Our basic idea is to determine $k_\alpha+k_\beta$ by comparing spin-relaxation measurements of alkali-metal atoms in \tHe\ and \fHe.
The spin-relaxation rate of the alkali-metal atoms due to \tHe\ is, at low polarization and low enough temperatures that the alkali-alkali spin-relaxation
rates can be ignored,
\begin{equation}
\tgA=^3\!k\dens{\tHe}=\left(\tkSR+k_\alpha+k_\beta\right)\dens{\tHe}
\end{equation}
where $k_{SR}$ is the relaxation produced by the spin-rotation interaction. 
The spin-relaxation rate due to
\fHe\ is simply
\begin{equation}
\fgA=\fkSR\dens{\fHe}
\end{equation}
since there is no spin-exchange for \fHe.  Thus we can use the relaxation of the alkali atoms in \fHe\ gas to isolate the spin-exchange
and spin-relaxation contributions. We argue below that the spin-relaxation rates for the two isotopes scale linearly with  the
collision velocities, so that
\begin{equation}
\tkSR=\sqrt{\mu_4\over \mu_3 }\fkSR \label{scale}
\end{equation}
where  $\mu$ is the reduced mass of the He-alkali pair.
This scaling should allow us to separate the spin-exchange and spin-rotation contributions to the alkali spin-relaxation rate:
\begin{equation}
k_\alpha+k_\beta= {^3k}-\sqrt{\mu_4\over \mu_3 }\fkSR\label{expt}
\end{equation}
Thus subtracting the scaled \fHe\ spin-relaxation rate from the \tHe\ spin-relaxation rate isolates the sum of the isotropic and anisotropic spin-exchange rates.

Experimentally, the challenge is to measure the alkali spin-relaxation rates carefully enough to preserve significance for the
subtraction in the numerator of Eq.~\ref{expt}.  The Rb-\tHe\ spin-exchange rate has now been measured by two different groups
\cite{Baranga98c,Chann02c} to be
$6.8\times10^{-20}$ cm$^3$/s.  The relaxation rates for Rb-He are unfortunately about 16-50 times bigger (depending on
temperature) than the spin-exchange rates.  Thus very high precision measurements would need to be made.

The situation is much better for potassium, where the measured efficiencies suggest a factor of 10 more favorable ratio of
spin-exchange to spin-relaxation rates.

We now turn to the scaling relation for the two isotopes.  The spin-rotation coupling $\gamma(R)$, $R$ being interatomic separation,
is inversely proportional to the reduced mass $\mu$ of the colliding pair.  This is because the rotation frequency of the atoms
about each other is
\begin{equation}
\boldbox{\omega}={\hbar {\bf N}\over \mu R^2}
\end{equation} 
which give rise to a Coriolis interaction
\begin{equation}
V_\omega=-\hbar\boldbox{\omega}\cdot{\bf L}
\end{equation}
where $\bf L$ is the electronic angular momentum\cite{Walker97b}.  The spin-rotation coupling then arises due to the response of the
electron to the effective magnetic field $B=\hbar\omega/(g_S\mu_B)$.   Thus one expects on very general grounds that
$\gamma(R)\propto {1/\mu}$.

The spin-relaxation rate coefficient is an average over the possible collision trajectories \cite{Walker89}
\begin{eqnarray}
k_{\rm SR}&=&\frac{8\pi\overline{v}\mu^2}{3\hbar^2}\int_0^\infty we^{-w}dw b^3 db\nonumber\\
&&\times \left|\int_{r_o}^{\infty}
\frac{\gamma(R)dR}{\sqrt{(1-b^2/R^2)-V(R)/w kT}}\right|^2,
\label{ssr}\end{eqnarray}
where $w$ is a dimensionless variable and $b$ the impact parameter of the collision.  $V(R)$ is the Rb-He potential, which should
be very insensitive to the mass of the He nucleus.  The inverse scaling of $\gamma$ with reduced mass cancels the $\mu^2$ factor in
front of the integrals, so that the mass-dependence of the spin-relaxation rate coefficient arises entirely from the relative
velocity factor $\overline{v}\propto 1/\sqrt{\mu}$.

\section{Experiment}

The K-He spin-relaxation measurements were made at Amersham Health using a 7.1 cm diameter spherical valved cell containing K metal
with a very small amount of Rb metal dissolved in it.  The Rb vapor density was  measured to be $2\pm0.4 \times 10^{-3}$ that of
the K.  The Rb atoms were polarized to typically 20\% polarization (parallel to a 20 G magnetic field) by optical pumping with a 60 W diode laser. The polarized Rb
atoms then polarized the K atoms by spin-exchange collisions.  A mechanical shutter periodically blocked the laser light to allow 
the alkali polarization to decay due to spin-relaxation. 

A
single-frequency tunable diode laser, operating at typically 3 nm or more from the potassium D1 line at 770 nm, was used to monitor
the spin-polarization of the alkali atoms by Faraday rotation.     The spin-relaxation transients were then analyzed to extract the
slowest decay mode of the relaxing atoms. This procedure was repeated a number of times as the pressure and composition of the cell
was varied.  Two decays were taken at each pressure, with different probe laser intensities.  A linear extrapolation to zero probe laser
intensity was performed to remove the effect of the probe laser (at most a 5\% correction).

Three gases were used for the experiments.  The ``\tHe'' gas was actually a 0.9922:0.0078 \tHe-N$_2$ mixture that was the standard
Amersham gas mixture.  Pure nitrogen gas was also used so that the nitrogen contribution to the \tHe\ relaxation could be corrected
for.  The third gas was $^4$He.  The cell was filled with the gas of interest at high pressure.  Immediately after filling with the
fresh gas, the alkali vapor pressure would suddenly drop, then slowly recover over the period of about an hour.  The drop in
pressure was presumably due to chemical reactions with impurities in the gases.  To vary the gas pressure, hot gas was pumped out
through the cell valve.  Since this was done with the cell hot, the gas density was determined from the pressure using the ideal gas
law at the 150$^\circ$C cell temperature.

\begin{figure}[htb]
\includegraphics[scale=0.58]{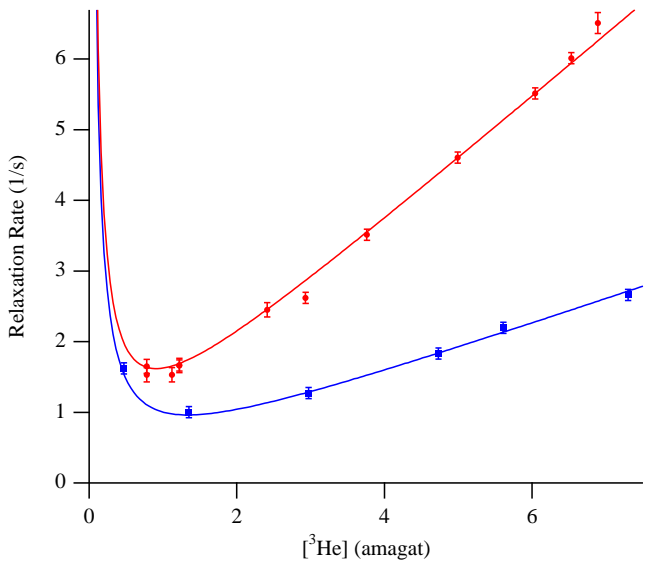}
\caption{Spin-relaxation data for K in two gas mixtures.  The top data is the \tHe-N$_2$ mixture, the bottom for pure
\fHe.}\label{data}
\end{figure}

The measured spin-relaxation decay rates are shown in Fig.~\ref{data}.  On a given day, the data vary smoothly with pressure;
however we found some systematic day-to-day changes that are outside the normal statistical fluctuations.  For example, the \tHe\
data points at 6.9 amagat and 2.9 amagat were taken on different days than most of the other data.  The size of these unexplained
fluctuations is about 4\%.

\section{Analysis} 

The data for the two gases were fit to the following function:
\begin{equation}
\Gamma=D_0\sqrt{\mu_3\over\mu_G} \left(\pi\over R\right)^2{{\rm 1 amagat}\over \dens{G}}+\Gamma_0+k\dens{G}
\end{equation}
with the first term representing diffusion, the second K-K relaxation, and the third spin-relaxation due to K-G collisions.  Based
on S. Kadlecek's thesis\cite{KadlecekThesis}, we expect $\Gamma_0<0.1$/s  at this temperature and this parameter was actually taken to be zero for the fit.
The data for both gases were fit simultaneously, assuming that the diffusion coefficient scales inversely with the square root of
the reduced mass $\mu$ of the K-G pair.  Thus only 3 parameters,  $D_0$, $k({\tHe})$, and $k({^4{\rm He}})$, were used to fit the
entire data set.  The results are:
\begin{equation}
\begin{array}{l}
D_0=0.91 \pm 0.04 \;\mbox{ cm$^2$/s} \\
k({\tHe})=0.89\pm0.04 \mbox{/s-amagat}\\
k({^4{\rm He}})=0.36\pm0.014 \mbox{/s-amagat}\\
\end{array}
\end{equation}
with the error bars reflecting the unexplained day-to-day fluctuations in the results.

At the 150$^\circ$C temperature, the K and Rb atoms are well into the regime where the spin-exchange rates between the alkali-metal atoms greatly exceed the
spin-relaxation rates for the atoms.  Thus the atoms should be well-described by a spin-temperature.  The presence of the Rb vapor
at a concentration of 1/500 slightly modifies the usual slowing down factor of 6 for a nuclear spin-3/2 atom like K to
$s=6+10.8/500=6.02$.  We also must account for a slight amount of Rb-He spin-relaxation, measured by Baranga {\it et al}
\cite{Baranga98c} to be 41.2/s-amagat for Rb\tHe, and, using the mass scaling, 36.1/s-amagat for $^4$He.  We therefore find
\begin{eqnarray}
\fkSR&=&s\times 0.36-36.1/500=2.10\;\mbox{/s-amagat}\nonumber\\
&=&7.8\times10^{-20}\;\mbox{cm$^3$/s}
\end{eqnarray}
and, using the mass scaling,
\begin{eqnarray}
\tkSR&=&1.14\fkSR=2.39\;\mbox{/s-amagat}\nonumber\\
&=&8.9\pm0.4\times10^{-20}\;\mbox{cm$^3$/s}
\end{eqnarray}
These are the first measurements of spin-relaxation  of K by He.

Since the \tHe\ gas is actually a mixture, a correction for N$_2$ must also be made.  From Ref.~\cite{KadlecekThesis}, and confirmed by a
measurement at 28 psig, we find that nitrogen contributes 1.24/s-amagat for the 0.78\% mixture used.  We therefore find for the
total K-\tHe\ spin-destruction rate coefficient (spin-exchange plus spin-rotation),
\begin{equation}
^3k=s\times 0.89-1.24-41.2/500=4.04\;\mbox{/s-amagat}\nonumber
\end{equation}
The spin-exchange contribution is therefore
\begin{eqnarray}
k_\alpha+k_\beta&=&^3\!k-\tkSR=1.65\;\mbox{/s-amagat}\nonumber\\
&=&6.1\pm0.7\times10^{-20} {\mbox{cm$^3$/s}}
\end{eqnarray}
The latest measurements of $k_{\rm SE}$ \cite{Babcock05c} give $k_\alpha-k_\beta/2=5.5\pm0.2\times10^{-20}$ cm$^3$/s.  Therefore the X-factor due to anisotropic
spin-exchange is
\begin{equation}
X_0={6.1\pm 0.7\over 5.5\pm 0.2}-1=0.11\pm0.13
\end{equation}
  This in turn implies that spin-exchange using K-\tHe\ collisions is fundamentally limited to a \tHe\ polarization of
\begin{equation}
P_{\rm max}={1\over 1+X_0}=0.90\pm0.11
\end{equation}
This result, though it does not rule out $P_{\rm max}=1$, is tantalizing since it suggests there may actually be a fundamental contribution to the X-factor.  Higher precision measurements of both the spin-exchange rate coefficient and the spin-relaxation measurements are needed to reach a definitive conclusion.

We can combine our \tHe\ spin-relaxation results with $k_{\rm SE}$ to obtain a spin-exchange efficiency
\be
\eta={k_{\rm SE}\over ^3k}={5.5\pm 0.2\over 15.0\pm 0.7}=0.37\pm 0.02
\ee
This is in slight disagreement (1.25 $\sigma$) with the Baranga {\it et al.} result \cite{Baranga98c} of 0.295$\pm$0.06, which was in turn found consistent with our previous observations of the efficiency of hybrid spin-exchange \cite{PhysRevLett.91.123003}.

We have presented in this paper a method for isolating the anisotropic hyperfine interaction from the much larger isotropic hyperfine interaction for alkali-metal atoms interacting with $^3$He.  This issue is not only of interest for fundamental reasons, but it has practical importance for maximizing the attainable polarization in spin-exchange optical pumping.  Considerable effort at NIST and Wisconsin \cite{Babcock06} has gone into trying to improve the wall-relaxation performance of \tHe\ spin-exchange.  The best polarization observed to date is 81\%.  If this value is approaching the fundamental limit for the process, there is little to be gained through further laborious wall studies.  On the other hand, if the limit is 95\% or higher, there is room to substantially improve the performance of spin-exchange pumped targets for applications such as neutron spin-filters \cite{Babcock09}, magnetic resonance imaging \cite{Holmes08}, and electron scattering \cite{Slifer08}.

\acknowledgements{
This work was supported by the National Science Foundation,  the Department of Energy, and Amersham Health.  The authors benefitted from discussions with T. Gentile and B. Driehuys.}

\bibliography{/Users/Thad_Walker/Research/thadbibtex/spinexchange}

\end{document}